\documentstyle[12pt]{article}
\normalbaselines
\begin{document}

\bibliographystyle{unsrt}
\vbox {\vspace{6mm}}

\begin{center}
{\bf WIGNER'S PROBLEM AND ALTERNATIVE COMMUTATION RELATIONS FOR 
QUANTUM MECHANICS}
\end{center}

\begin{center}
V. I. MAN'KO,$^\star$\footnote{On leave from Lebedev Physical
Institute, Moscow, Russia.} G. MARMO,$^\dagger $ 
E. C. G. SUDARSHAN,$^\ddagger $ and F. ZACCARIA$^\dagger $
\end{center}

\begin{center}
$^\star ${\it Osservatorio Astronomico di Capodimonte\\
Via Moiariello 16, 80131 Napoli, Italy}
\end{center}  

\begin{center}
$^\dagger $ {\it Dipartimento di Scienze Fisiche,
Universit\`a di Napoli ``Federico II''\\
 and\\
Istituto Nazionale di Fisica Nucleare, Sezione di Napoli\\
Mostra d'Oltremare, Pad.19 - 80125 Napoli, Italy}
\end{center}

\begin{center}
$^\ddagger ${\it Physics Department, Center for Particle Physics\\
University of Texas, Austin, Texas, 78712 USA}
\end{center}

\begin{abstract}
It is shown, that for quantum systems the vectorfield associated 
with the equations of motion may admit alternative Hamiltonian descriptions,
both in the Schr\"odinger and Heisenberg pictures. We illustrate these
ambiguities in terms of simple examples.
\end{abstract}

\section{Introduction}

\noindent

Wigner studied the following problem:\cite{wig} to what extent do the
equations of motion determine the quantum mechanical commutation relations?
He found that the commutation relations are not determined uniquely by
the equations of motion even if the form of the Hamiltonian is fixed.
The question was reexamined by Green,\cite{green} where he found a 
large family of possible commutation (and anticommutation) relations, 
which are now well-known as parastatistics. They are associated with the 
realization of the compact algebras $~B_n=o\,(2n+1)~$ and the 
noncompact algebra $~C_n =sp\,(2n,\,R)\,.$\cite{Ryan}

The problem of ambiguities of quasidistribution functions
was discussed\cite{wigok} and this problem is intrinsically related to
the ambiguities in the choice of commutation relations.
The analogous problem has been addressed also for systems in classical 
mechanics. In this context, it is known\cite{marmo,dodma,hojman} that
the equations of motion may be obtained using alternative Hamiltonians 
and alternative symplectic structures (Poisson brackets) as well as
alternative Lagrangians.

From this analysis, it is clear that if we start with a given vectorfield 
(dynamics) on the space of states, we may find several additional
alternative structures defined on the carrier space which are compatible
(i.e., they are preserved) with the dynamical evolution.
Thus, the same carrier space may be endowed with several structures
as long as all of them are compatible with the dynamics.
In quantum mechanics, when the states are described by wave functions,
the carrier space is usually endowed with a vector space structure
and with a Hermitean structure (i.e., it is considered as a Hilbert space),
which are preserved by the dynamical evolution. 

In this paper, we would like to take up this last aspect and consider the 
problem raised by Wigner also in the Schr\"odinger picture. It is clear, 
that by thinking of the Hilbert space structure
as given by a vector space structure plus a real and imaginary parts 
of the Hermitean structure, we may have alternative descriptions by
giving, on the same space, alternative real and (or) imaginary parts of the
Hermitean structure.\cite{boris} 
To make clear which structures are involved, we shall work mainly on
finite dimensional Hilbert space and eventually give some examples
in infinite dimensional vector spaces.

To show how we may benefit of our knowledge from classical
mechanics, we will start with the simplest possible example:
the one-level system.

\section{One-Level Systems}

\noindent

We start with a one-level quantum system to show that the Schr\"odinger
equation for it gives rise to a classical-like Hamiltonian dynamics for a
one-dimensional harmonic oscillator.

In facts, the Schr\"odinger equation 
\begin{equation}\label{a1}
i\hbar \,\frac {\partial \Psi (t)}{\partial t}=H\Psi (t)\,,
\end{equation}
where $~\Psi (t)$ is the wave function of the one-level system, is 
described by the Hamiltonian operator $~H$. This operator is a Hermitean
1$\times $1--matrix and it means that the Hamiltonian is simply a 
$~c$-number, which is real. If one introduces the two real variables 
$~q\,(t)~$ and $~p\,(t)~$ as real and imaginary parts of the wave function
\begin{equation}\label{a2}
\Psi (t)=\frac {1}{\sqrt 2}\,[\,q\,(t)+i\,p\,(t)\,]~,
\end{equation}
the Schr\"odinger equation acquires the form
\begin{equation}\label{a3}
\dot q=\frac {H}{\hbar }\,p\,;~~~~~~
\dot p=-\frac {H}{\hbar }\,q\,.
\end{equation}
Then, setting $~\omega =H/\hbar \,,$ one can rewrite Eqs.~(\ref{a3}) as
\begin{equation}\label{a5}
\ddot q + \omega ^2 q=0\,,
\end{equation}
which represents the equation of motion for the one-mode harmonic
oscillator with frequency $~\omega ~$ and with the Hamiltonian
\begin{equation}\label{a6}
H=\omega \left (\frac {p^2}{2}+\frac {q^2}{2}\right )\,.
\end{equation}
We may call this system a classical-like system (the remark that quantum 
wave equation may be rewritten as classical-like equation was done
in paper\cite{str}\,).

It is known,\cite{marmo,dodma} that there exists an infinite number of 
variational formulations for one-dimensional classical systems. Is
it possible to use such alternative Hamiltonian descriptions to construct 
alternative ``quantum descriptions'' for our system~(\ref{a1})\,? 
To answer properly this question, we need to examine which structure 
are involved in the descriptions of quantum systems.
              
\section{Structures for Quantum Descriptions}

\noindent

We consider a finite dimensional Hilbert space $~{\cal H}~$ 
of dimension $~N\,.$ By using Dirac notation, $~|\Psi \rangle \,$ is 
{\it state} vector, $~|\Psi \rangle \in {\cal H}\,$ and
$~\lambda |\Psi \rangle \,$ represent the same physical state for any
$~\lambda \in {\hbox {\bf C}}\,.$

{\it Observables} are associated with Hermitean matrices $~A=A^\dagger \,.$
Equivalently, one could describe observables in terms of  bilinear
functions (the summation over repeated indices is assumed)
\begin{equation}\label{b1}        
f_A(\psi )=\psi _k^* A_{kl}\,\psi _l\,,
\end{equation} 
which are known as expectation values.
Below, we set Planck constant equal to unity.
With functions, we associate infinitesimal transformations.

Usually the change in the wave function associated with an
infinitesimal generator $~f_A~$ is given by
\begin{equation}\label{b3}
 \delta _{f_A}\psi_k =-i\,\varepsilon \,\frac {\partial f_A}{\partial
\psi _k^*}\,;~~~~~~~~ \delta_{f_A}\psi_k^*=-i\,\varepsilon \,
\frac {\partial f_A}{\partial \psi _k}\,.
\end{equation}
In this way, the transformaion associated with the identity matrix gives
$$ \delta _{f_1}\psi _k=-i\,\varepsilon \,\psi_k\,;~~~~~~~\delta _{f_1} 
\psi _k^*=i\,\varepsilon \,\psi_k^*\,.$$
Using this standard assumption, the change in some other function
$~f_B~$ induced by $~f_A~$ is given by
\begin{eqnarray}\label{b4}
\delta _{f_A}f_B &=
 & \frac {\partial f_B}{\partial \psi _k}
\,\varepsilon \,\delta _{f_A}\psi _k  
+ \frac {\partial f_B}{\partial \psi _k^*}  
\,\varepsilon \,\delta _{f_A}\psi _k^*\nonumber\\
 &= &-i\,\varepsilon \left [\frac {\partial f_B}{\partial \psi _k}
\,\frac {\partial f_A}{\partial \psi _k^*}
-\frac {\partial f_A}{\partial \psi _k^*}\,\frac {\partial f_B}
{\partial \psi _k}\right ]\nonumber\\
&=&i\,\varepsilon \,(\psi _l^*B_{lk}A_{km}\psi _m
-\psi _l^*A_{lk}B_{km}\psi _m ~)\nonumber\\
&=&i\,\varepsilon \langle \psi|(AB - BA)|\psi \rangle \,.  
\end{eqnarray}
We can summarize by saying that $X_A\,,$ defined by
\begin{equation}\label{b6}
\delta_{f_A}=\varepsilon X_A=(\delta_{f_A} \psi_k)
\,\frac {\partial}{\partial \psi_k} +
(\delta_{f_A} \psi_k^*)
\,\frac {\partial}{\partial \psi_k^*}\,,
\end{equation}
is the vectorfield associated with $~f_A~$ and the
Poisson bracket is defined by
\begin{equation}\label{b7}
X_Af_B=i\,\{f_A,\,f_B\}=i\,f_{[A,B]}\,.
\end{equation}
By considering the vectorfield associated with the Hamiltonian,
$~f_H(\psi)=\langle \psi|H|\psi \rangle \,,$ we find a vectorfield 
$~X_H\,,$ whose action on a function $~f~$ is given by
\begin{equation}\label{b8}
X_Hf=i\,\left (\psi_k^*H^{kl}\frac {\partial f}{\partial \psi^*_l}\right )
-i\,\left (\psi_k^*H^{kl}\frac {\partial f}{\partial \psi^*_l}\right )^*\,.
\end{equation}
We can write now the associated equations of motion in the Schr\"odinger 
picture
\begin{equation}\label{b9}
\frac {d\psi_k}{dt}=-i\,\frac {\partial f_H}{\partial \psi^*_k}\,;
~~~~~~\frac {d\psi^*_k}{dt}=i\,\frac {\partial f_H}{\partial \psi_k}\,,
\end{equation}
or, in what may be called the {\it Ehrenfest picture},
\begin{equation}\label{b10}
L_{X_H}f_B=\frac {d}{dt}f_B=i\,\{f_H,\,f_B\}\,,
\end{equation}
where $~L_{X_H}~$ stays for the Lie derivative along $~X_H\,.$

From here, we derive also the equations of motion in the Heisenberg
picture
\begin{equation}\label{b11}
i\,\frac {d}{dt}A=[A,\,H]\,.
\end{equation}
We have found that equations of motion can be written on the Hilbert space,
on quadratic functions (expectation values), and on operators (observables),
associated with the Schr\"odinger, Ehrenfest, and Heisenberg pictures,
respectively.

It is now appropriate to consider which structures play a role in the
various pictures we are examining.

From the Poisson bracket on quadratic functions~(\ref{b7}), we derive a 
Poisson bracket on linear functions by using the derivation property, 
namely,
\begin{equation}\label{b12}
if_{[A,B]}=i\,\{f_{A},\,f_{B}\}=\frac{1}{2}\,\left (\frac {\partial f_A}
{\partial x_i}\,\{x_i,\,x_j\}\,\frac {\partial f_B}{\partial x_j} 
-\frac {\partial f_B}{\partial x_i}\,\{x_i,\,x_j\}
\,\frac {\partial f_A}{\partial x_j}\right )\,,
\end{equation}
and solving with respect to $~\{x_i,\,x_j\}\,.$ Here we have used collective
coordinates $~x_i~$ with $~i=1,\,\ldots ,\,2n~$ instead of $~(\psi^*_k,\,
\psi_k)~$ with $~i=1,\,\ldots ,\,n\,.$

If the Poisson bracket is expressed as
\begin{equation}\label{b13}
\{x_i,\,x_j\}=c_{ij}\,,
\end{equation}
then we can construct a symplectic structure
\begin{equation}\label{b14}
\omega = \omega^{ij}\,dx_i\wedge dx_j
\end{equation}
with $~\omega ^{im}c_{mj}=\delta ^i_j\,.$

We recall, that if on a real finite-dimensional vector space $~V~$ 
we have an endomorphism $~J\,:\,V\rightarrow V~$ with the property
$~J^2=-{\hbox {\bf 1}}\,,$ then $~V~$ has even dimension, say $~2N\,,$ 
and because of $~J\,(Jx)=J^{2}x=-x\,,$ i.e., the eigenvalues of $~J~$ 
are pure imaginary, $~J~$ is called a complex structure on $~V\,.$

By using $~\omega \,,$ we can now construct a real valued bilinear 
function $~s$
\begin{equation}\label{b15}
s\,(x,\,y)=\omega \,(Jx,\,y)\,.
\end{equation}
When $~s~$ is positive definite and $~s\,(x,\,y)=s\,(y,\,x)\,,$ we say 
that $~J~$ and $~\omega ~$ are compatible. In this case, we can define 
the following bilinear function $~h~$ on $~V$
\begin{equation}\label{b16}
 h\,(x,\,y)=\omega \,(Jx,\,y)+(J\omega)(x,\,y)\,,
\end{equation}
or by considering $~V~$ as a complex vector space
\begin{equation}\label{b30}
h\,(x,\,y)=s\,(x,\,y)+i\,\omega \,(x,\,y)\,.
\end{equation} 
This bilinear function defines a Hermitean scalar product, giving $~V~$
the structure of a Hilbert space.

Now a quantum system, as determined, for instance, in~(\ref{b9}), will be 
a vectorfield on the Hilbert space preserving the Hermitean structure 
$~h~$ on the complex vector space $~V\,.$
Clearly it has to preserve both the positive definite bilinear function
$~s\,:\,V\times V\rightarrow \mbox {\bf R}\,,$ and the skew-symmetric
bilinear function $~\omega \,:\,V\times V\rightarrow \mbox {\bf R}\,,$ 
where $~V~$ is thought as a real vector space. Associated to $~s~$ and 
$~\omega \,,$ it is known that maps $~\hat s\,:\,V\rightarrow V^*\,$
and $~\hat \omega \,:\,V\rightarrow V^*\,$ are isomorphisms.
It follows, that $~J=\hat s^{-1}\hat \omega \,:\,V\rightarrow V~$ is 
also preserved by the vector field.

Our dynamical vectorfield thought as an element of a Lie algebra preserves:

i) a symplectic structure, i.e., it is an element of $~sp\,(2N,\,R)\,;$

ii) a positive definite scalar product, i.e., it is an element of 
$~so\,(2N,\,R)\,;$

iii) a complex structure, i.e., it is an element of $~gl\,(n,\,\mbox 
{\bf C})\,.$

As a matter of fact, the intersection of any two of such algebras determines
the algebra of the unitary group $~u\,(n,\,\mbox {\bf C})\,,$ coherently 
with the interpretation of our vectorfield as a quantum system.

We can restate our findings in the language of differential geometry.
A linear vectorfield $~X\,,$ associated with equations of motion on the
real vector space $~V\,,$ represents a quantum system if it admits
invariant symplectic structures $~\omega \,,$
i.e., $~L_X\omega =0\,$ (again $~L_X~$ stays for the Lie derivative), and
invariant complex structure $~J\,,$ i.e., $~L_{X}J=0\,,$ which are 
compatible (we recall, that compatibility means that 
$~\omega \,(Jx_1,\,x_2)~$ is a real valued symmetric positive definite 
bilinear function). With any such pair $~(\omega ,\,J)\,,$ we construct 
a Hilbert space structure on $~V\,.$ Our starting vectorfield $~X~$ becomes 
a ``quantum system'' with respect to the constructed Hilbert space structure.

To make contact with the Heisenberg version of the same statement, we need
a preliminary results connecting the Poisson bracket on the vector space
with the Poisson bracket on quadratic functions and as well as
the relation connecting these brackets with Lie products on square matrices.

We have: any Poisson bracket on quadratic functions determines a Poisson
bracket on linear functions with the property 
$~\{x,\,y\}\in \mbox {\bf R}\,.$ The converse statement is obvious.

We consider the derivation property of Poisson brackets and then have
\begin{equation}\label{b17}
\{f_A,\,f_B\}=\frac{1}{2}\,\left (\frac {\partial f_A}{\partial x_i} 
\,\{x_i,\,x_j\}\,\frac {\partial f_B}{\partial x_j} 
-\frac {\partial f_B}{\partial x_i}\,\{x_i,\,x_j\}\,\frac {\partial f_A}
{\partial x_j}\right )\,.
\end{equation}
Since by assumption $~\{f_A,\,f_B\}~$ is also a quadratic function, both
$~\partial f_A/\partial x_i~$ and $~\partial f_B/\partial x_i~$ 
are linear and therefore  $~\{x_i,\,x_j\}~$ must be a numerical matrix, 
say $~\{x_i,\,x_j\}=c_{ij}\,.$
By using the arbitrariness of $~A~$ and $~B\,,$ we can solve for 
$~||c_{ij}||\,.$ We can also show, that the correspondence between square 
matrices and quadratic functions is a Lie algebra isomorphism.

With any matrix $~A=||A^{ij}||\,,$ we associate a function
 $~f_A=x_{i}A^{ij}x_{j}\,.$

Vice versa to any bilinear function $~f\,,$ we associate a matrix 
$~A_f~$ with 
$$A_f^{~ij}=\frac {\partial ^2f}{\partial x_i\,\partial x_j}\,.$$
If we are given Poisson brackets on quadratic functions, we can define a 
product on matrices, by setting
\begin{equation}\label{b18}
[A, B]_{\{f_A,\,f_B\}}^{~ij}=\frac {\partial ^2}{\partial x_i\,
\partial x_j}\,\{f_A,\,f_B\}\,.
\end{equation}
This product defines a Lie algebra structure on square matrices.

Vice versa, if we have a Lie product on matrices, say 
$$[A,\,B]=C\,,$$
we define a Poisson bracket on quadratic functions by setting
$$\{f_{A},\,f_{B}\}=f_C\,.$$
The Jacobi identity on the algebra of matrices is equivalent to 
the Jacobi identity for the Poisson bracket.

\vskip .5cm

These various brackets allow us to write the 
dynamics in the various pictures, namely:

a) Schr\"odinger picture
\begin{equation}\label{b19}
i\,\frac {dx}{dt}=\{f_{H},\,x\}=\frac {\partial f_H}
{\partial x_i}\,\{x_{i},\,x\}\,;
\end{equation}

b) Ehrenfest picture
\begin{equation}\label{b20}
i\,\frac {df_A}{dt}=\{f_H,\,f_A\}\,;
\end{equation}

c) Heisenberg picture
\begin{equation}\label{b21}
i\,\frac {d}{dt}\,A=[A,\,H]\,.
\end{equation}

Previous  relations between brackets on the various carrier
spaces link the descriptions of equations of
motion in the different pictures.

\section{Alternative Commutation Relations}

\noindent

Going back to our vectorfield $~X~$ on the real vector space $~V\,,$ 
we say that $~X~$ admits alternative quantum descriptions, if there is 
more than one compatible pair $~(\omega ,\,J)~$ preserved by $~X\,.$
From what we have said, there will be a corresponding alternative quantum
description in the Ehrenfest picture and in the Heisenberg picture. 
Are we likely to find alternative descriptions for a given vectorfield
$~X\,,$ which admits the compatible pair $~(\omega ,\,J)\,?$ 

We can actually show that there are quite a few alternative descriptions
for a ``quantum system'' on a finite dimensional Hilbert space $~{\cal H}~$
of complex dimension $~n\,.$ If 
$$ X_A=A^i_jx_i\frac {\partial }{\partial x_j}$$
is a quantum system with respect to the Poisson bracket
$$\{x_i,\,x_j\}_C=c_{ij}$$
and $~C~$ is such that
$$C=||c_{ij}||\,,$$ 
a compatible complex structure $~J~$ will be a matrix, such that
$~[J,\,A]=0~$ and $~CJ=s~$ will be a matrix representation of a
positive definite scalar product.
We can be more explicit concerning the relations among matrices associated
with the various structures.
We assume that our dynamics is Hamiltonian, i.e., we have
\begin{equation}\label{b22}
\frac {dx}{dt}=L_{X_A}x=\{f_{H},\,x\}\,,
\end{equation}
which gives the following relation among the matrix $~C~$ associated
with the Poisson brackets, the symmetric matrix $~H~$ associated with
the quadratic function $~f_H\,,$ and the matrix $~A~$ representing the 
vectorfield $~X_A\,:$
\begin{equation}\label{b23}
 1. ~~A=H\,C\,.
\end{equation}
We also have the compatibility conditions
\begin{eqnarray*}
2.&~~&A\,J=J\,A\,,~~~~~~~J^2=-\mbox {\bf 1}\,;\nonumber\\
3.&~~&C\,J=s\,.\nonumber
\end{eqnarray*}
Consider now any invertible transformation $~T~$ on the vector space 
$~V\,,$ which is a symmetry for $~A\,,$ i.e. $~T^{-1}AT=A\,.$ By 
applying it to any one of the previous relations, we find:
\begin{eqnarray}\label{b25}
1_{T}.&~~&A=T^{-1}H\,^t(T^{-1})\,^tT\,C\,T=H_TC_T\,;\\
&&\nonumber\\
2_{T}.&~~&T^{-1}J\,T\,T^{-1}A\,T=T^{-1}\,A\,T\,T^{-1}\,J\,T\,,~~
~~~J_T\,A=A\,J_T\,,~~~~~J_T^2=-\mbox {\bf 1}\,;\nonumber\\
3_{T}.&~~&^tT\,C\,T\,T^{-1}J\,T= {}^tT\,s\,T\,,
~~~~~~C_T\,J_T=s_T\,.\nonumber
\end{eqnarray}
Therefore any symmetry for $~A~$ which is not a unitary transformation will
provide an alternative pair $~(\omega _T,\,J_T)~$ for $~X_A\,.$

In finite dimensions, the symmetry group for a generic matrix $~A~$ is
generated by matrices $~A^0,\,A^1,\,\ldots ,\,A^{2n}\,,$ i.e., powers 
of $~A~$ completely generate the symmetry group for $~A\,.$

From the decomposition $~A=HC\,,$ simple reflection shows that odd
powers still satisfy the decomposition into a symmetric matrix times the 
skewsymmetric $~C\,,$ while for even powers this is not true.\cite{rubano} 
Thus, even powers generate transformations which are not unitary, therefore
they will give rise to alternative quantun descriptions.

In the coming subsections, we show how alternative descriptions in the present
picture will give rise to alternative structures in a different picture.
We first derive some preliminary relations.

\subsection{Alternative Lie products on matrices associated with various 
Poisson brackets}

\noindent

Starting from $~\{x_i,\,x_j\}_C=c_{ij}\,,$ we investigate what kind of Lie 
product we are going to get on square matrices.

As we said in previous sections, we shall find
\begin{equation}\label{c1}
[A_{f},\,A_{g}]_C=A_{\{f,g\}_C}\,.
\end{equation}
By carrying on the computation, in symbolic notation, we find
\begin{equation}\label{c2}
[B_1,\,B_2]_C=B_1C\,B_2-B_2C\,B_1\,.
\end{equation}
 We should stress that the right-hand side stays for 
$$B_1^{il}C_{lm}B_2^{mj}-B_2^{il}C_{lm}B_1^{mj}\,,$$
i.e., the Lie product is defined on matrices having the transformation 
properties of $~(0,\,2)$-tensors instead of $~(1,\,1)$-tensors, as we 
would have for linear transformations.

\subsection{New Lie products on the space of linear transformations}

\noindent

Inspired by the previous Lie product, obtained from quadratic functions,
we can define now a new Lie product in the space of linear transformations.

If $~A,\,B,\,K \in \mbox {Lin}\,(V,\,V)\,,$ we define a new associative 
product on linear transformations by setting
\begin{equation}\label{c3}
A\cdot_K\,B=A\,e^{\lambda K}B\,.
\end{equation}
It is not difficult to show that this product is associative and 
distributive. With it, we associate a Lie product by setting
\begin{equation}\label{c4}
[A,\,B]_K=A\cdot_K\,B-B\cdot_K\,A=A\,e^{\lambda K}B-B\,e^{\lambda K}A\,.
\end{equation}
We can easily find that
\begin{equation}\label{c5}
[A,\,B\cdot_KC]=[A,\,B]_K\cdot_KC+B\cdot_K[A,\,B]_K\,,
\end{equation}
i.e., this product defines derivations with respect to the new associative
product we have defined. The Jacobi identity is also easily obtained.

We consider now the map
$$A\rightarrow F_K(A)=e^{\lambda K/2}\,A\,e^{\lambda K/2}\,,$$
we see that
\begin{equation}\label{c6} 
F_K(A)\,F_K(B)=F_K(A\cdot_KB)\,,
\end{equation}
i.e., we have a homomorphism of associative algebras. We derive from it the
Lie algebra homomorphism
\begin{equation}\label{c7}
[F_K(A),\,F_K(B)\,]=F_K(\,[A,\,B]_K)\,.
\end{equation}
The variety of Lie products we have defined can be put to work to find
alternative descriptions for our dynamical vectorfield.

\subsection{Alternative quantum descriptions}

\noindent

We start with a vectorfield $~X~$ preserving an Hermitean structure $~h\,,$ 
also denoted for simplicity in the bra-ket notation
$$h\,(\psi_1,\,\psi_2)=\langle \psi_1|\psi_2\rangle_h\,,$$
or directly  $~\langle \psi_1|\psi_2\rangle~$ without the suffix $~h\,,$ 
when no confusion arises. With this Hermitean structure, we associate  
Poisson brackets on the vector space (defined via the imaginary part of 
$~h\,).$ Our system will be Hamiltonian with Hamiltonian function $~f_H\,.$

Now we start with a product $~A\cdot_KB~$ on the vector space
of linear transformations and define quadratic function
\begin{equation}\label{c8}
f_{A,K}(\psi)=\langle \psi |e^{\lambda K/2}\,A\,
e^{\lambda K/2}|\psi \rangle _h
=\langle \psi |F_K(A)|\psi \rangle _h\,.
\end{equation}
If we denote by
$$\langle \psi_1|\psi_2\rangle _K=\langle \psi_1|e^{\lambda K/2}
\,e^{\lambda K/2}|\psi_2\rangle _h\,,$$
we get
\begin{equation}\label{c9}
f_{A,K}=\langle \psi |A|\psi \rangle _K\,.
\end{equation}
Thus, we have defined a new Hilbert space structure on our vector space
by the use of an operator $~K~$ which is symmetric with respect to the
starting scalar product associated with $~h\,.$ Of course, it is still 
symmetric with respect to the new product, as it can be easily verified.

Quadratic functions, we have defined, with the new scalar product on operators
will define new Poisson brackets, i.e., we are ready to use all our previous
relations between different pictures. The starting vectorfield $~X~$
will admit all these new structures as alternative ones, if the operator 
$~K~$ is a constant of the motion for $~X\,.$ Thus, the present construction 
associates a family of alternative descriptions with any constant of the
motion.

Until now, we have considered finite dimensional Hilbert spaces and 
matrices acting on them. We shall now consider the example Wigner was 
examining in his paper.\cite{wig} 

\section{The Harmonic Oscillator}

\noindent

We consider the quantum oscillator and set frequency and mass to be
unity along with Planck constant.
In terms of complex amplitude operators $a$ (annihilation operator) and
$a^{\dag}$ (creation operator), we have
\begin{eqnarray}\label{d1}
\dot a+i\,a&=&0\,;\\
\dot a^\dagger-i\,a^\dagger &=&0\,.\nonumber
\end{eqnarray}
The commutation relation
$$a\,a^\dagger-a^\dagger a=1$$
allows us to write the equation of motion in Hamiltonian form with
$$ H=a^\dagger a+\frac {1}{2}\,.$$
All this is standard. We recall that in the standard treatment, we have the
following relations
\begin{eqnarray*}
a|n\rangle &=&\sqrt n\,|n-1\rangle \,;\nonumber\\
a^\dagger|n\rangle &=&\sqrt {n+1}\,|n+1\rangle\,.\nonumber
\end{eqnarray*}
Now we consider a new Lie product on our operators
\begin{equation}\label{d2}
[a,\,a^\dagger ]_K=a\,e^{\lambda K(a^\dagger a)}\,a^\dagger-
a^\dagger e^{\lambda K(a^\dagger a)}\,a\,.
\end{equation}
By using 
$$ a\,K(a^\dagger a)=K(a^\dagger a+1)\,a$$
and 
$$a^\dagger K(a^\dagger a)=K(a^\dagger a-1)\,a^\dagger \,, $$
we find
\begin{equation}\label{d3}
[a,\,a^\dagger ]_K=(e^{\lambda K(\hat n+1)}-e^{\lambda K(\hat n-1)})
\,\hat n+e^{\lambda K(\hat n+1)}\,,
\end{equation}
where we have set $~\hat n=a^\dagger a\,.$
              
Now we write the equations of motion with the new commutator, namely
\begin{equation}\label{d4}
[\tilde H,\,a]_K=\tilde H\,e^{\lambda K}\,a-a\,e^{\lambda K}\,
\tilde H=-i\,a\,;
\end{equation}
$$
[\tilde H,\,a^\dagger ]_K=\tilde H\,e^{\lambda K}\,a^\dagger
-a^\dagger e^{\lambda K}\,\tilde H=i\,a^\dagger \,,$$
to find
\begin{equation}\label{d5}
\tilde H\,e^{\lambda K}=H=a^\dagger a+\frac {1}{2}\,.
\end{equation}
Thus, any $~K~$ and $~\tilde H ~$ satisfying
$$\tilde H\,e^{\lambda K}= H=a^\dagger a+\frac {1}{2}$$
will produce an alternative description.
For instance, setting
$$\tilde H=\frac {\sinh \,\lambda \hat n}{\sinh \lambda}$$
requires
$$e^{\lambda K}=\frac {(\hat n+1/2)\,\sinh {\lambda}}
{\sinh \,\lambda \hat n}\,.$$

It is interesting to solve for the ``standard commutation relation,'' 
i.e., to solve for $~K(\hat n)~$ the equation
$$ a\,e^{\lambda K}a^\dagger -a^\dagger e^{\lambda K}a=1\,.$$
In other terms, we are investigating on the uniqueness
of the function $~K(\hat n)\,.$ Indeed, we know that $~K=0~$ gives the
standard result.

In the Fock basis, our equation becomes
$$(n+1)\,e^{\lambda K(n+1)}-n\,e^{\lambda K(n-1)}=1\,.$$
This is equivalent to the recurrence relation
$$e^{\lambda K(n+1)}-\frac {n}{n+1}\,e^{\lambda K(n-1)}=
\frac {1}{n+1}\,.$$
For odd integers, we have choosing $~K(1)=0\,,$
$$ e^{\lambda K(2s+1)}=1\,;~~~~~s=0,\,1,\,2,\,\ldots \,.$$
But for even integers, we have the one-parameter family of values
\begin{eqnarray*}
e^{\lambda K(0)}&=&1+\varepsilon \,;\nonumber\\
e^{\lambda K(2)}&=&1+\frac {1}{2}\,\varepsilon \nonumber\\
e^{\lambda K(2s)}&=&\frac {(2s-1)\,(2s-3)\,\cdots \,1}{2s\,(2s-2)\,\cdots
\,2}\,\varepsilon \,.\nonumber
\end{eqnarray*}
The action of $~\lambda K(\hat n)~$ on the Hilbert subspace of odd states
is different from the zero operator. This deformation of the usual commutation
relation is closely related with possible different behaviour of even and
odd coherent states and commutation property with respect to the parity
operator. It should be remarked, that this property had been noticed by 
Wigner,\cite{wig} where the discussion was done not in terms of creation 
and annihilation operators but in terms of position and momentum operators.

To make contact with some previous work where consequences for correlation
function were considered, we present here an equivalent way to change
the commutation relation. This approach gives rise again to a new scalar
product (a new Hilbert space structure), introducing some kind of local
inhomogeneous elastic deformation.

We consider the following transformation on the basis operators
\begin{eqnarray*}
a&\rightarrow &a\,f\,(\hat n)=A\,;\nonumber\\
a^\dagger &\rightarrow &f\,(\hat n)\,a^\dagger =A^\dagger \,.\nonumber
\end{eqnarray*}
Clearly, this transformation does not change previous equations of 
motion (it is a symmetry for them because $~f(\hat n)~$ is a constant
of the motion). Indeed, we get
\begin{eqnarray*}
\dot A+i\,A&=&0\,;\nonumber\\
\dot A^\dagger -i\,A^\dagger &=&0\,.\nonumber
\end{eqnarray*} 
Commutation relations for $~A,\,A^\dagger ~$ are given by
$$ A\,A^\dagger -A^\dagger A=\varphi \,(F^{-1}(\hat N)\,)\,,$$
where
 $$F\,(n)=f^{2}(n)\,n\,;~~~~~n=F^{-1}(N)\,,$$
and $~\varphi \,(x)~$ is related to $~f\,(x)~$ by the equation
$$\varphi \,(x)=(x+1)\,f^2(x+1)-x\,f^2(x)\,.$$
It is however possible to define a new scalar product in the vector space
of states by constructing them in the usual manner. The vacuum state is
the same for both $~a~$ and $~A\,,$ i.e., 
$$a|0\rangle =0$$ 
and 
$$ A|0\rangle =0\,,$$ 
and we construct
$$|n\rangle =\frac {(a^\dagger )^n}{\sqrt {n!}}\,|0\rangle $$
and
$$|N\rangle =\frac {(A^\dagger )^n}{\sqrt {n!}}\,|0\rangle $$
and define two scalar products
$$\langle n|m\rangle _{h_1}=\delta_{nm}$$
and 
$$\langle N|M\rangle _{h_2}=\delta_{NM}\,.$$
Now we find
$$\langle M|(A\,A^\dagger -A^\dagger A)\,|N\rangle _{h_2}=\delta_{NM}$$
similar to
$$\langle m|(a\,a^\dagger -a^\dagger a)\,|n\rangle _{h_1}=\delta_{nm}\,.$$

It is clear now that both scalar products and commutation relations are 
compatible with the equations of motion and provide allternative
descriptions. This new product can be given in the form of our 
generalized associative product by considering an operator $~K\,,$ such
that
$$A=a\,f(\hat n)=e^{\lambda K/2}\,a\,e^{\lambda K/2}\,,$$
and we find the following relation between the function describing the 
nonlinearity and the operator
$$f\,(\hat n)=\exp \left \{\frac {1}{2}\,\lambda \,[K(\hat n - 1) 
+K(\hat n)]\right \}\,.$$

{\it Remark}.~~~This relation cannot be satisfied by any $~K\,,$ if 
$~f(n)~$ has zeroes. If the function $~f\,(n)~$ becomes zero for 
physical (nonnegative) integer values of $~n_0+1\,$ then the operator 
$~A~$ vanishes on $~|n_0\rangle \,.$ So the
vacuum or the states $~|n\rangle \,,$ with $~0<n<n_0\,,$ are no longer 
cyclic with respect to the algebra generated by $~A~$ and 
$~A^\dagger \,.$ The original vector space decomposes into a direct sum.

So instead of an infinite set of basis vectors starting with the vacuum
we have only a finite number of discrete energy states. This is an
important case to study in practice, since the 
harmonic-oscillator-potential approximation is valid only for low 
excitations beyond which the energy levels are no longer equally spaced 
or even discrete energy bound states. Models of diatomics molecules and 
of an ion in a Paul trap are immediately relevant examples.

\section{Conclusions}

\noindent

In this paper, we have studied the interrelations of the vectorfield
(dynamics) defining the ``equations of motion,'' the kinematic 
characterization of the Poisson bracket and the possible Hamiltonians 
that generate the motion. Some aspects related to the discussed problems 
are available.\cite{f,solimeno} For convenience in presentation
and to avoid lengthy investigations of domains of definition of operators,
we have restricted our attention to systems with a finite number of 
states, to which one may associate a (classical-like) system with the same
number of degrees of freedom as the number of states.

On the finite dimensional real vector space, carrying a quantum dynamical
vectorfield, we have identified a complex structure $~J\,,$ Poisson brackets
associated with a matrix $~C~$ and a positive definite symmetric matrix
$~s=C\,J\,,$ defining a scalar product. The inverse of the Poisson
bracket defines a symplectic structure $~\omega \,.$ By using $~s~$ and
$~\omega \,,$ we construct an Hermitean (structure) product;  vice versa
the real and imaginary parts of an Hermitean product can be identified
with $~s~$ and $~\omega \,,$ respectively. With every Hermitean product
which is preserved by the evolution of our starting system, we associate
a possible quantum description. All these have been called alternative
quantum descriptions. 

By considering the evolution in terms of expectation values (quadratic
functions), which we have called Ehrenfest picture, and the associated
one in terms of operators (matrices), which provides the dynamics in
the Heisenberg picture, we have found that alternative quantum
descriptions have a counterpart also in these pictures. In particular,
when we consider vectorfields on the space of operators, we are back
to the formalism used by Wigner in his paper.\cite{wig} Here it is shown 
that one may construct several alternative associated products on the
space of operators compatible with the dynamical evolution.

We hope, we have made clear to what extent the equations of motion do not 
determine the commutation relations.

Another aspect, which however requires a deeper understanding, has to do
with the implications for the physical interpretation of these
alternative quantum descriptions. Briefly, this could be stated as the
problem of identification of the physical variables: which variables
are amenable to direct and immediate measurement? The answer to
this question must be based on the experience. We are not meerely
talking about the degree of precision that we can obtain but the question
of which are the variables that a given observer finds as natural
variables to measure. Different observers could use different commutation
relations on the same space of operators or different scalar products on
the same space of states. Which of them should be considered as
physically equivalent observers?
For instance, one of important possible physical consequences of the
obtained results is that might exist quantum vibrations of quadratures
(momentum and position), some respecting Heisenberg relations and others
respecting different uncertainty relations. These questions should be further
analyzed and they would go beyond the scope of the present paper.

Finally, we should mention that alternative Hamiltonian descriptions
show up in the theory of completely integrable systems.\cite{mag1,mag2} 
Our alternative quantum descriptions do imply that in the previous meaning
quantum systems are completely integrable systems.\cite{cir,cas,mavi}

\section*{Acknowledgements}

\noindent

V. I. M. thanks Osservatorio Astronomico di Capodimonte for hospitality
and Russian Foundation for Basic Research under Project No. 17222 for
partial support.

\end{document}